\documentclass{aastex701}
\submitjournal{RNAAS}

\usepackage{amsmath}
\usepackage{bm}
\usepackage{cleveref}

\begin{document}

\title{The Pulsar Radial Acceleration Relation}

\author{Tariq Yasin}
\email{tariq.yasin@physics.ox.ac.uk}
\affiliation{Astrophysics, University of Oxford, Denys Wilkinson Building, Keble Road, Oxford, OX1 3RH, UK}

\author{Harry Desmond}
\email{}
\affiliation{Institute of Cosmology \& Gravitation, University of Portsmouth, Dennis Sciama Building, Portsmouth, PO1 3FX, UK}

\begin{abstract}
The radial acceleration relation (RAR) links observed and baryonic
accelerations, and is best established in rotation curves of late-type galaxies.
Pulsar timing, which measures line-of-sight (LOS) differential accelerations between the Sun and pulsars, provides a novel probe of this relation, including along directions outside the Galactic disc.  By combining these pulsar differential accelerations with the acceleration at the Sun, we test whether current pulsar timing data carry information
on a vector generalisation of the RAR,
${\bm g}_{\rm obs}=\nu(|{\bm g}_{\rm bar}|){\bm g}_{\rm bar}$.  Comparing the measured SPARC RAR (generalised to 3D) to 26 binary-system pulsars with literature accelerations, we find a reduced $\chi^2$ of 3.58, compared with 10.86 for Newtonian baryonic gravity alone. However, setting all accelerations to that of the Sun gives a reduced $\chi^2$ of 3.75, showing that this vector RAR test is dominated by the Solar acceleration with current data.
\end{abstract}

\section{Introduction}

The radial acceleration relation (RAR) is a tight empirical correlation
between the inferred centripetal acceleration $g_{\rm obs}$ and the
baryonic gravitational acceleration $g_{\rm bar}$, established primarily in disc
galaxies~\citep{McGaugh2016,Varasteanu2025}.  Its near-zero intrinsic scatter, simple
functional form, and lack of secondary correlations
~\citep{Desmond2023b,Desmond2023,StiskalekDesmond2023} have been viewed as evidence
for non-Newtonian dynamics~\citep[e.g.][]{Famaey2012} or as a consequence of
coupled baryon--halo assembly in $\Lambda$CDM~\citep[e.g.][]{Paranjape2021}; here we treat it agnostically as simply an empirical mapping. Whether the RAR phenomenology extends beyond the radial dynamics of late-type galaxies measured from rotation curves remains unsettled \citep[e.g.][]{Lelli2017,Brouwer2021,Freundlich2022,Mistele2024,Julio2025}. Pulsar timing
provides a new probe of Galactic accelerations outside the Galactic disc: binary orbital-period derivatives measure line-of-sight (LOS) accelerations through the acceleration-induced drift of the Doppler shift~\citep{Phillips2021,Chakrabarti2021,Moran2024,Donlon2024}. In this note, we use these measurements to test a phenomenological RAR rather than infer a full Galactic acceleration field~\citep{Donlon2024,Donlon2025} or test a theory-specific model.  We assume a local vector relation
${\bm g}_{\rm obs}=\nu(|{\bm g}_{\rm bar}|){\bm g}_{\rm bar}$ which we project onto the Sun--pulsar sightline.  This tests whether pulsar timing, one of very few non-rotation-curve acceleration probes, is already informative about this striking relation.

\section{Data and model}

We use the v3 catalogue of~\citet{Donlon2025}, which has 26 independent LOS
accelerations inferred from orbital-period derivatives of pulsars in
binary systems after Shklovskii and relativistic corrections. The catalogue also contains a sample of spin-only millisecond-pulsar accelerations (inferred with empirical magnetic-braking corrections). However we did not find that sample informative for our test due to the larger uncertainties, and so we report results for the binary sample only.

We compute the baryonic field from an analytic approximation to the
photometric Milky Way census of~\citet{BlandHawthorn2016}.  We represent their thin
disc, thick disc, cold gas and bulge/inner-disc components by three
Miyamoto--Nagai discs with
$(M,a,b)=(3.5\times10^{10},2.6,0.3)$, $(6\times10^9,2.0,0.9)$ and
$(1.0\times10^{10},4.5,0.08)$, denoting mass and radial/vertical scale
lengths in $M_\odot$ and kpc, plus a Hernquist bulge with mass
$1.5\times10^{10}M_\odot$ and scale radius $0.6\,{\rm kpc}$.
  The gas
disc normalisation reproduces their quoted local gas surface density,
$\Sigma_g(R_0)\simeq13\,M_\odot{\rm pc}^{-2}$.  Observed pulsar positions are transformed to Galactocentric coordinates using
$R_0=8.20\pm0.09\,{\rm kpc}$~\citep{McMillan2017} and
$z_\odot=20.8\,{\rm pc}$~\citep{BennettBovy2019}.  Distance
uncertainties are propagated approximately by re-evaluating the field at $d+\sigma_d$
and $d-\sigma_d$ and taking half the difference.

Pulsar timing measures the differential LOS acceleration of the pulsar
relative to the Sun,
\begin{equation}
  a_{\rm LOS}^{\rm diff}
  =({\bm g}_{\rm psr}-{\bm g}_\odot)\cdot\hat{\bm\ell},
\end{equation}
where ${\bm g}_{\rm psr}$ and ${\bm g}_\odot$ are Galactocentric
accelerations and $\hat{\bm\ell}$ points from the Sun to the pulsar. Timing gives no information on components orthogonal to the
Sun--pulsar sightline.  We therefore compute $a_{\rm LOS}^{\rm obs}$, the pulsar's Galactocentric
acceleration projected onto the Sun--pulsar sightline:
\begin{equation}
  a_{\rm LOS}^{\rm obs}
  =a_{\rm LOS}^{\rm diff}+{\bm g}_\odot\cdot\hat{\bm\ell}
  ={\bm g}_{\rm psr}\cdot\hat{\bm\ell}.
\end{equation}
We adopt the Solar acceleration from~\citet{McMillan2017} $a_{\odot,R}=-v_0^2/R_0=-2.15\times10^{-10}\,{\rm m\,s}^{-2}$ (corresponding to circular velocity $v_0=233.1\pm3.0\,{\rm km\,s}^{-1}$). The vertical Solar acceleration is
evaluated from the baryonic model and is 1.7\% of the centripetal term. As our primary output we plot $a_{\rm LOS}^{\rm obs}/({\bm g}_{\rm bar}\cdot\hat{\bm\ell})$, the ratio of the observed to baryonic Galactocentric accelerations projected along the line of sight. 

The vector RAR prediction for $a_{\rm LOS}^{\rm obs}$ is
\begin{equation}
  a_{{\rm LOS},i}^{\rm pred}
  =\nu(|{\bm g}_{{\rm bar},i}|)
   ({\bm g}_{{\rm bar},i}\cdot\hat{\bm\ell}_i),
\end{equation}
with $\nu=1$ for Newtonian baryon-only gravity. We adopt the Simple interpolating function as constrained by galaxy dynamics in~\citet{Desmond2023},
\begin{equation}
  \nu(x)=\frac{1}{2}+\sqrt{\frac{1}{4}+\frac{a_0}{x}},
  \quad a_0=1.19\times10^{-10}\,{\rm m\,s}^{-2}.
\end{equation}

We compare observed and predicted accelerations with
$\chi^2_{\rm LOS}=\sum_i(a_{{\rm LOS},i}^{\rm obs}
-a_{{\rm LOS},i}^{\rm pred})^2/\sigma_i^2$, where $\sigma_i$ includes
timing, Solar-correction, and distance-propagated baryonic-field
uncertainties. 

Converting the pulsar differential timing acceleration to a Galactocentric acceleration requires combining it with the empirical Solar centripetal acceleration. With the present sample, the median Solar acceleration fraction,
$f_\odot \equiv |a_{\odot,\rm LOS}|/(|a_{\rm LOS}^{\rm diff}|+|a_{\odot,\rm LOS}|)$,
is 0.73 for the binary sample.  We therefore additionally conduct a Solar-acceleration-only test ($a_{\rm LOS}^{\rm diff}=0$) to isolate the effect of the pulsar-specific accelerations.

\section{Results and discussion}

\begin{figure*}
\centering
\includegraphics[width=\textwidth]{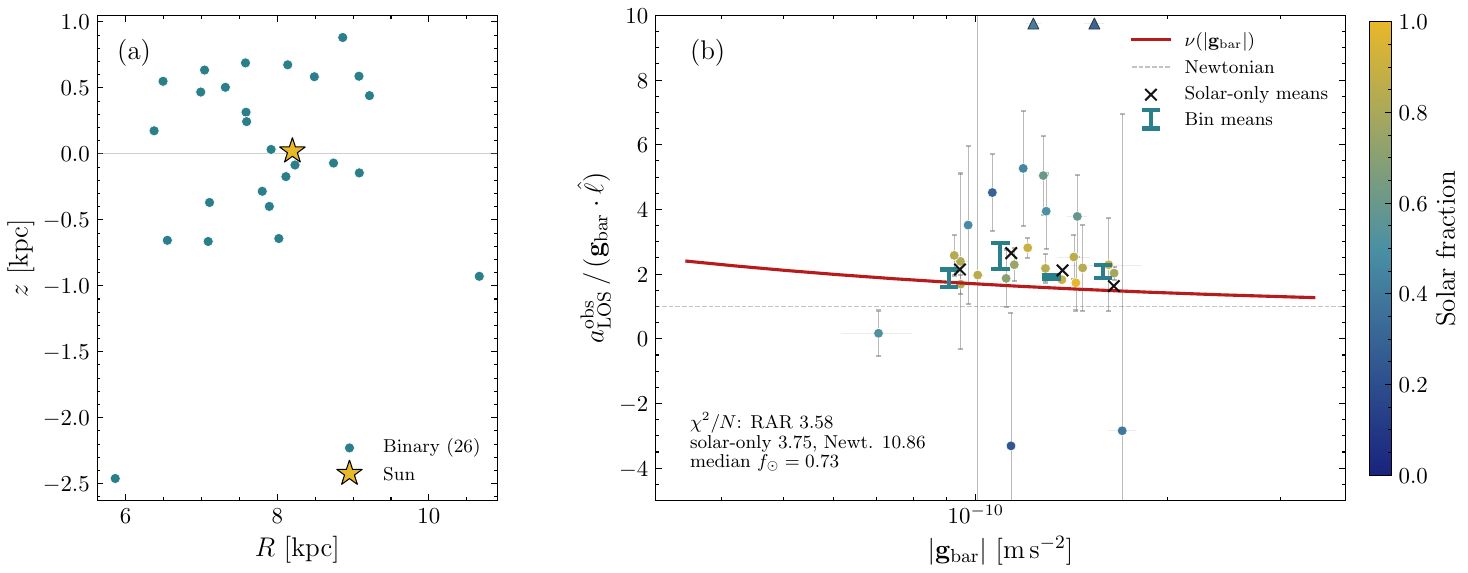}
\caption{(a) Galactocentric positions of the binary pulsar sample (circles).  (b) Boost ratio
$a_{\rm LOS}^{\rm obs}/({\bm g}_{\rm bar}\cdot\hat{\bm\ell})$ versus
$|{\bm g}_{\rm bar}|$.  The red curve is the Simple RAR fit to galaxy rotation curves of \citet{Desmond2023b} and the grey
line is Newtonian.  Points are coloured by the fraction of acceleration
magnitude that is from the Solar acceleration. The cyan error bars show inverse-variance-weighted means in equally spaced bins in $\log_{10} g_{\rm bar}$ (only bins with two or more pulsars are shown). The black crosses show the
Solar-only bin means (i.e. setting the measured differential
acceleration to zero), offset horizontally for visual clarity. Triangles mark points clipped beyond the plotting range.}
\label{fig:combined}
\end{figure*}

For the binary sample, the projected RAR gives
$\chi^2_{\rm LOS}/N=3.58$ (for $N=26$ binaries), compared with 10.86 for Newtonian baryonic
gravity.  Neither value gives a satisfactory absolute goodness of fit, but
the projected RAR is substantially closer to the data.  The plot in
\cref{fig:combined} shows this in ratio form, but the scatter is
large and the sampled range in $|{\bm g}_{\rm bar}|$ is only about
0.8 dex around $a_0$.  These data therefore cannot distinguish non-radial RAR
extensions or theory-specific Galactic acceleration fields. The Solar-only null (setting
$a_{\rm LOS}^{\rm diff}=0$ while retaining each sightline's
Solar projection) gives $\chi^2_{\rm LOS}/N=3.75$, close to
the real binary value.  This shows that the result is determined in large part by the Solar acceleration rather than by
the pulsar-specific differential accelerations.
Using the Milky Way baryonic potential of~\citet{McMillan2017} instead gives $\chi^2/N=3.94$ for the projected RAR, 10.74 for Newtonian
baryons, and 4.16 for the Solar-only comparison, demonstrating qualitative insensitivity to this systematic.


In conclusion, pulsar timing offers a direct probe of acceleration-based scaling relations, but the current binary sample is not yet significantly more informative than the Solar acceleration contribution alone. More precise binary acceleration measurements, larger samples, improved distances, and sightlines for which the Solar acceleration is a smaller fraction of the Galactocentric acceleration are required for more informative tests.  The most useful future objects will be binary pulsars at lower $|{\bm g}_{\rm bar}|$, larger $|z|$, or Galactic longitudes for which the Solar centripetal acceleration projects weakly onto the Sun--pulsar sightline.

\vspace{40mm}

\bibliography{references}{}
\bibliographystyle{aasjournalv7}

\end{document}